\documentclass{jfm}
\usepackage{graphicx}
\usepackage{epstopdf, epsfig}
\usepackage{mathtools}
\usepackage{amsmath}
\usepackage{amssymb}
\usepackage{xcolor}

\newcommand{\be}{\begin{eqnarray}}
\newcommand{\ee}{\end{eqnarray}}
\newcommand{\bse}{\begin{subequations}}
\newcommand{\ese}{\end{subequations}}

\newcommand{\bpm}{\begin{pmatrix}}
\newcommand{\epm}{\end{pmatrix}}

\newcommand{\pt}{\partial}

\newcommand{\lap}{\nabla^2}

\newcommand{\vv}[1]{\boldsymbol{#1}}
\newcommand{\grad}{\boldsymbol{\nabla}}

\newcommand{\Iota}{\mathrm{i}}

\usepackage{xcolor}
\newcommand{\rev}[1]{{\color{black} #1}}

\DeclareGraphicsExtensions{.png,.pdf}

\shorttitle{Linearly forced rotating flows}
\shortauthor{R. Supekar, V. Heinonen, K. J. Burns and J. Dunkel}

\title{Linearly~forced~fluid~flow~on~a~rotating~sphere}
\author{Rohit Supekar\aff{1,2},
Vili Heinonen\aff{2},
Keaton J.\ Burns\aff{2},
J\"{o}rn Dunkel\aff{2}\corresp{\email{dunkel@mit.edu}}}

\affiliation{\aff{1} Department of Mechanical Engineering, Massachusetts Institute of Technology, 77~Massachusetts Avenue, Cambridge, MA 02139, USA
\aff{2} Department of Mathematics, Massachusetts Institute of Technology, 77~Massachusetts~Avenue, Cambridge, MA 02139, USA}

\begin{document}

\maketitle

\begin{abstract}
 \rev{We investigate generalized Navier--Stokes (GNS) equations
    that couple nonlinear advection with a generic linear instability}. This analytically tractable minimal model for fluid flows driven by internal active stresses has recently been shown to permit exact solutions on a stationary 2D sphere. 
\rev{Here, we extend the analysis to linearly driven flows on rotating spheres}. We derive exact solutions of the GNS equations corresponding to time-independent zonal jets and superposed westward-propagating Rossby waves, \rev{qualitatively similar to those seen in planetary atmospheres}.  
Direct numerical simulations with large rotation rates obtain statistically stationary states close to these exact solutions. 
The measured phase speeds of waves in the GNS simulations agree with analytical predictions for Rossby waves. 
\end{abstract}


\section{Introduction}
Turbulence is often described as the last unsolved problem in classical physics \citep{Falkovich2006}. 
In recent years, considerable progress has been made in the modelling of stationary turbulence, which requires a driving force to continually balance kinetic energy losses due to viscous dissipation~\citep{Frisch1995}.  Theoretical and computational studies of turbulence phenomena typically focus on external driving provided by a random forcing  \citep{Boffetta2012a},  boundary forcing \citep{Grossmann2016} or Kolmogorov forcing \citep{Lucas2014}. A fundamentally different class of internal driving mechanisms, less widely explored in the turbulence literature so far, is based on linear instabilities~\citep{Rothman1989,Tribelsky1996,Sukoriansky1999,Rosales2005,Slomka2017a,Slomka2018,Linkmann2019}.  The profound mathematical differences between external and internal driving were emphasized by~\cite{Arnold1991} in the context of classical dynamical systems described by ordinary differential equations.  Specifically, he contrasted the externally forced Kolmogorov  hydrodynamic system with the internally forced Lorenz system, the latter providing a simplified model of atmospheric convection \citep{Lorenz1963}. From the broader fluid-mechanical perspective, Arnold's analysis raises the interesting question of how internally driven flows behave in rotating frames like the atmospheres of planets or stars.

\par
To gain insight into this problem, we investigate an analytically tractable minimal model for linearly forced quasi-2D flow on a rotating sphere. The underlying  generalized Navier-Stokes (GNS) equations describe internally driven flows through  higher-order hyperviscosity-like terms in the stress tensor~\citep{Beresnev1993,Slomka2017a}, and the associated GNS triad dynamics is structurally similar to the Lorenz system \citep{Slomka2018}. GNS-type models have been studied previously as effective phenomenological descriptions for seismic wave propagation \citep{Beresnev1993, Tribelsky1996}, magnetohydrodynamic flows \citep{Vasil2015} and active fluids~\citep{Slomka2017b,Slomka2017a,James2018}. A key difference compared with scale-free  classical turbulence is that GNS flows can  exhibit characteristic spatial and temporal scales that reflect the internal forcing mechanisms.

\par
\begin{figure}
	\centering
	\includegraphics[width=0.9\textwidth]{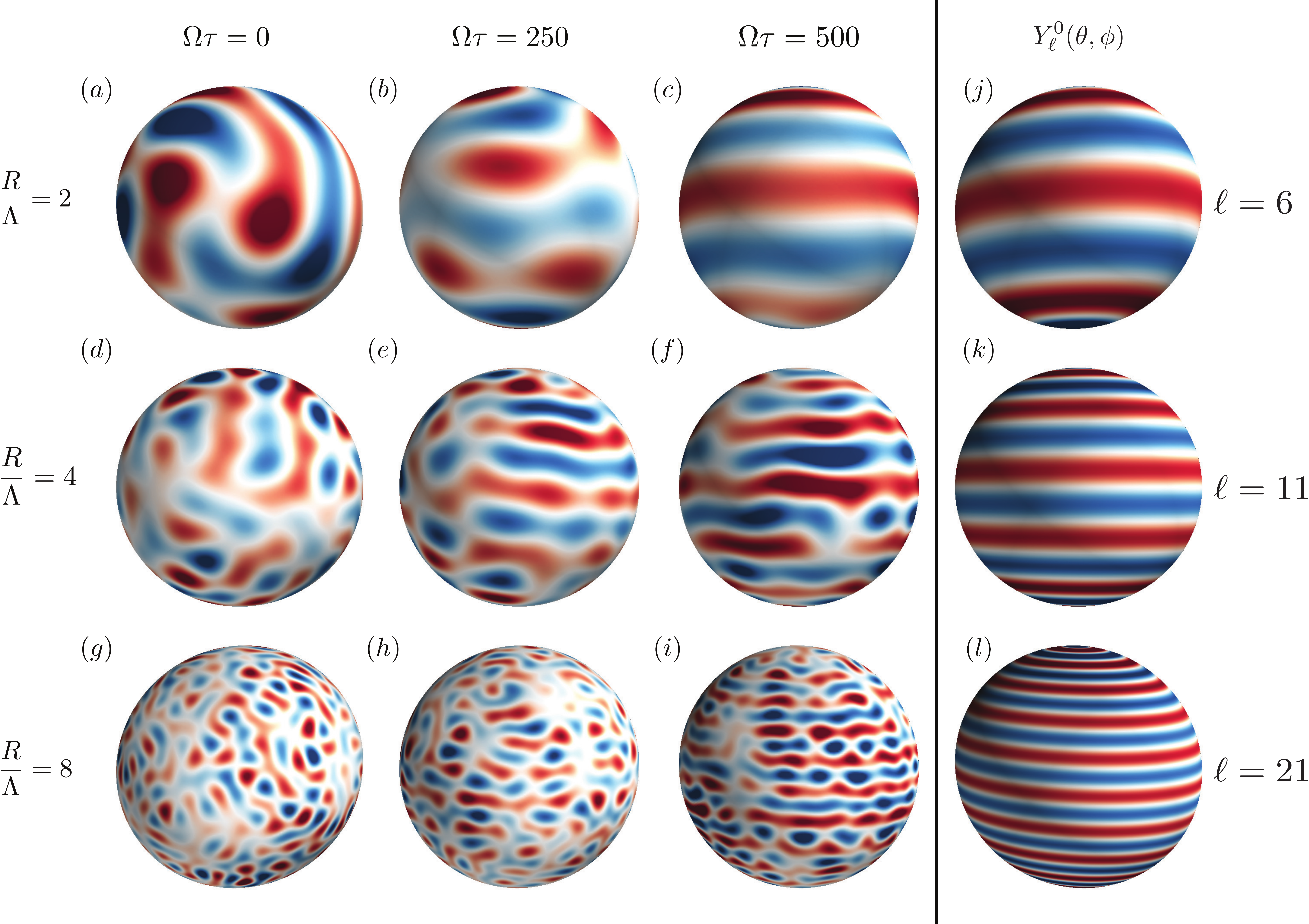}
	\caption{Statistically stationary states of the normalized vorticity $\zeta \tau$ from  simulations ($a-i$) for $\kappa \Lambda = 1$ become more zonal (or banded) as the rotation rate $\Omega \tau$ increases (see also supplementary movie 1). 
	At the highest rotation rate $\Omega \tau = 500$, the width of the alternating zonal jets is determined by the parameter $R/\Lambda$ that represents the ratio of the radius of the sphere and the diameter of the vortices forced by the GNS dynamics. 
	The main characteristics of these flow patterns at high rotation rate are captured by spherical harmonics $Y_\ell^0(\theta, \phi)$ that solve the dynamical equations. Matching the length scale $R/\Lambda$ gives $\ell = 6$ $(j)$, $\ell = 11$ $(k)$, and $\ell = 21$ $(l)$ for $R/\Lambda=2$, 4, and 8, respectively.
	}
	\label{Fig:Patterns}
\end{figure}

Remarkably, the minimal GNS model studied below permits nontrivial analytical solutions. Exact stationary solutions reported previously include 3D Beltrami flows~\citep{Slomka2017a} and 2D vortex lattices~\citep{Slomka2017b}. Furthermore, \cite{Mickelin2018} recently explored GNS flows on 2D curved surfaces and constructed stationary solutions for the case of a non-rotating sphere. Here, we generalize their work by deriving exact time-dependent solutions for GNS flows on rotating spheres, and by comparing them with direct numerical simulations (figure \ref{Fig:Patterns}). We shall see that these exact GNS solutions correspond to Rossby waves propagating along alternating zonal jets, qualitatively similar to the large-scale flow patterns seen in planetary atmospheres~\citep{Heimpel2005,Schneider2008}. 

\par
\rev{Our study complements recent work which showed that  non-equilibrium approaches can provide analytical insights into the dynamics of planetary flows~\citep{Delplace2017} and atmospheres \citep{Marston2012}. In view of the recent successful application of phenomenological GNS models to active fluids~\citep{Dunkel2013a,Slomka2017a}, the results below} can also help advance the understanding of active matter propagation on curved surfaces~\citep{Sanchez2012,Zhang2016,Henkes2018,Nitschke2019} and in rotating frames \citep{Lowen2019a}.

\section{Generalized Navier-Stokes model for linearly driven flow}
\label{Sec:Theory}

After briefly reviewing the GNS equations in \S \ref{Sec:GNSIntro}, we derive the corresponding vorticity-stream function formulation on a rotating sphere in \S \ref{Sec:GNSrotating}. 

\subsection{Planar geometry} \label{Sec:GNSIntro}
The GNS equations for an incompressible fluid velocity field $\vv{v}(\vv{x}, t)$ with pressure field $p(\vv{x}, t)$ read \citep{Slomka2017a, Slomka2017b}
\begin{subequations}
\label{Eq:GNSFlat}
\be
\grad \cdot \vv{v}  &=& 0, \\
\pt_t \vv{v} + (\vv{v} \cdot \grad) \vv{v} &=& - \grad p + \grad \cdot \vv{\sigma},
\noindent
\ee
where the higher-order stress tensor  
\be
\vv{\sigma} = \left( \Gamma_0 - \Gamma_2 \nabla^2 + \Gamma_4 \nabla^4 \right)   \left[ \grad \vv{v} + (\grad \vv{v})^\top\right]
\label{Eq:StressTensor}
\ee
\end{subequations}
accounts for both viscous damping and linear internal forcing. Transforming to Fourier space, the divergence of the stress tensor gives the dispersion relation 
\be
\xi(k) = -k^2 (\Gamma_0 + \Gamma_2 k^2 + \Gamma_4 k^4) 
\label{Eqn:GrowthFlat}
\ee
where $k$ is the magnitude of the wave vector $\vv{k}$. Fixing hyper-viscosity parameters $\Gamma_0>0$, $\Gamma_4 >0$ and $\Gamma_2 < - 2 \sqrt{\Gamma_0 \Gamma_4}$, the growth rate $\xi(k)$ is positive between the two real roots $k_-$ and $k_+$. Hence, Fourier modes in the active band $k \in (k_-,k_+)$ are linearly unstable, corresponding to active energy injection into the fluid. The distance between the neutral modes $k_\pm$ defines the active bandwidth $\kappa = k_+ - k_-$.
\par 
Unstable bands are a universal feature of stress tensors exhibiting  positive dispersion $\xi(k)>0$ for some $k$. Polynomial GNS models of the type \eqref{Eq:GNSFlat} were first studied in the context of seismic wave propagation \citep{Beresnev1993, Tribelsky1996} and can also capture essential statistical properties of dense microbial suspensions~\citep{Dunkel2013a,Slomka2017a}. Since non-polynomial dispersion relation produce qualitatively similar flows~\citep{Slomka2018,Linkmann2019}, we focus here on stress tensors of the generic polynomial form~\eqref{Eq:StressTensor}.  
\par
Exact steady-state solutions of \eqref{Eq:GNSFlat}, corresponding to \lq zero-viscosity\rq{} states, can be written as superpositions of modes $\vv{k}$ with $|\vv{k}| = k_+$ or $|\vv{k}| = k_-$ \citep{Slomka2017a}. Simulations of~\eqref{Eq:GNSFlat} with random initial conditions converge to statistically stationary states with highly dynamical vortical patterns that have a characteristic diameter $\sim\Lambda=\pi/k_*$, where $k_*$ is the most unstable wavenumber, corresponding to the maximum of $\xi(k)$~\citep{Slomka2018,SlomkaTownsend2018}. Inverse energy transport in 2D can bias the dominant vortex length scale towards larger values $\sim \pi/k_-$ \citep{James2018}.

\subsection{On a rotating sphere}
\label{Sec:GNSrotating}
We generalize planar 2D GNS dynamics \eqref{Eq:GNSFlat} to a sphere with radius $R$ rotating at rate $\Omega$. To this end, we adopt a co-rotating spherical coordinate system $(\theta, \phi)$ where $\theta$ is the co-latitude and $\phi$ is the longitude. Following \cite{Mickelin2018}, we find the rotating GNS equations in vorticity-stream function form
\bse
\be
\nabla^2 \psi  &=& -\zeta,\\ 
\qquad\quad
\p_t \zeta + J(\psi, \zeta) &=& F(\lap + 4K) (\lap + 2K) \zeta + 2 \Omega K \p_\phi \psi,
\label{Eqn:GNSRotating}
\ee
\ese

where $\zeta$ is the vorticity in the rotating frame, and $K = 1/R^2$ denotes the Gaussian curvature of the sphere. The active stress operator $F$ has the polynomial form
\be
F(x) = \Gamma_0 - \Gamma_2 x + \Gamma_4 x^2.
\ee

The Laplacian $\nabla^2$ on the sphere is defined by $
\nabla^2 = K \left(\cot \theta \pt_\theta + \pt_\theta^2 + (\sin \theta)^{-2} \pt_\phi^2 \right)
$ and $
J(\psi, \zeta) = 
K (\sin \theta)^{-1} \left( \p_\phi \psi  \p_\theta \zeta - \p_\theta \psi \p_\phi \zeta \right)
$ \rev{ is the determinant of the Jacobian of the mapping $(R \phi \sin \theta,R\theta) \mapsto (\psi,\zeta) $ from the tangent space of the sphere to the vectors $(\psi,\zeta)$.}
The velocity components can be recovered from the stream function $\psi$ by $
(v_\phi, v_\theta) = \left( -\pt_\theta \psi /R, \pt_\phi \psi/ R \sin \theta \right).$ In the non-rotating limit $\Omega \rightarrow 0$, \eqref{Eqn:GNSRotating} reduces to the model studied by \cite{Mickelin2018}. 
We note that \eqref{Eqn:GNSRotating} is an internally forced extension of the unforced barotropic vorticity equation $\p_t \zeta + J(\psi, \zeta) =   2 \Omega K \p_\phi \psi$ which has been widely studied in earth science since the pioneering work of \cite{Charney1950}. Below we will show that the GNS model~\eqref{Eqn:GNSRotating} is analytically tractable, permitting exact traveling wave solutions that are close to the complex flow states observed in simulations. 

\begin{figure}
    \centering
    \includegraphics[scale=0.8]{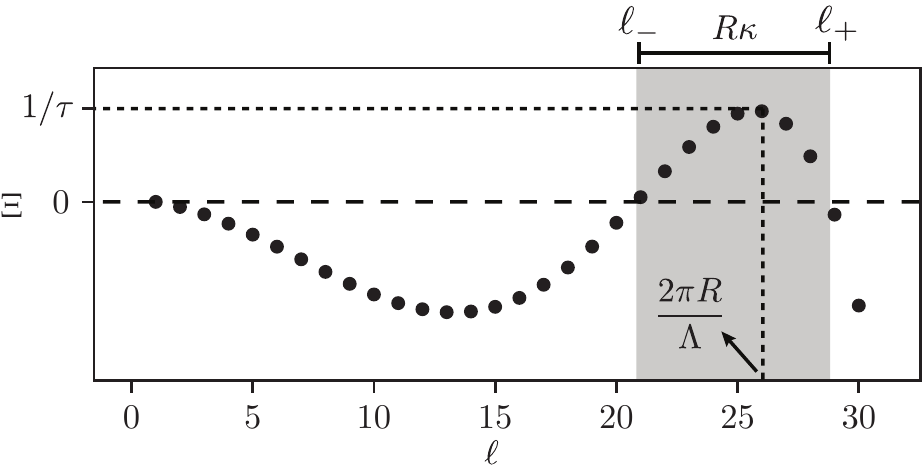}
    \caption{The growth rate $\Xi$ of spherical harmonic modes $Y_\ell^m (\theta, \phi)$ in \eqref{Eqn:GrowthRateSphere} plotted as a function of the wavenumber $\ell$. The parameters used to make this plot are \rev{$((\tau/R^2)\Gamma_0, (\tau/R^4) \Gamma_2, (\tau/R^6) \Gamma_4) \simeq (1.43 \times 10^{-2}, -4.86 \times 10^{-5}, 3.72\times 10^{-8})$ which correspond to $R/\Lambda = 8$ and $\kappa \Lambda = 1$.} The grey region indicates the active bandwidth where $\Xi > 0$ and energy is injected. $\Lambda$ is the diameter of the vortices forced by the mode with the maximum growth rate $1/\tau$, and $\kappa$ is the active bandwidth i.e. $\kappa = (\ell_+ - \ell_-)/R$ where $\Xi(\ell_\pm) = 0$.}
    \label{fig:DispRelation}
\end{figure}

\subsubsection{Dimensionless parameters}
We assess the linear behaviour of \eqref{Eqn:GNSRotating} using spherical harmonics $Y_\ell^m(\theta, \phi)$, the eigenfunctions of the Laplacian operator on the sphere. With $\delta = K(\ell (\ell+1) - 4)$, the linear growth rate of a spherical harmonic mode due to $F$ is 
\be
\Xi(\ell) = - \left(\delta + 2K \right) F(-\delta) = - \left(\delta + 2K \right) \left(\Gamma_0 + \Gamma_2 \delta + \Gamma_4 \delta^2 \right)
\label{Eqn:GrowthRateSphere}
\ee
which is the spherical analog of \eqref{Eqn:GrowthFlat}.  Using this relation, characteristic length and time scales, $\Lambda$ and $\tau$, for vortices forced by the GNS dynamics, along with the bandwidth of the forcing $\kappa$, can be expressed in terms of $\Gamma_0, \Gamma_2, \Gamma_4$ and $R$ (figure \ref{fig:DispRelation} \rev{and appendix \ref{App:Scales}}). We use these scales to define the essential dimensionless parameters: 
 $R/\Lambda$ is the ratio between the radius of the sphere and the characteristic vortex scale, $\kappa \Lambda$ compares the forcing bandwidth to the characteristic vortex scale, and $\Omega \tau$ is dimensionless rotation rate. \rev{We note that since $\Xi(\ell=1)=0$, the GNS dynamics do not force the $\ell = 1$ mode which ensures that the total angular momentum is conserved (see appendix \ref{App:AngularMomentum}).}  
Finally, we define the Rossby number in terms of the characteristic flow speed $\mathcal{U} = \Lambda/\tau$ and the dominant length scale $\mathcal{L} = \Lambda$ as
\be
Ro = \frac{ \mathcal{U}}{\Omega \mathcal{L}} = \frac{1}{\Omega \tau}. 
\label{Eqn:RossbyNumber}
\ee

\begin{figure}
	\centering
	\includegraphics[width=0.9\textwidth]{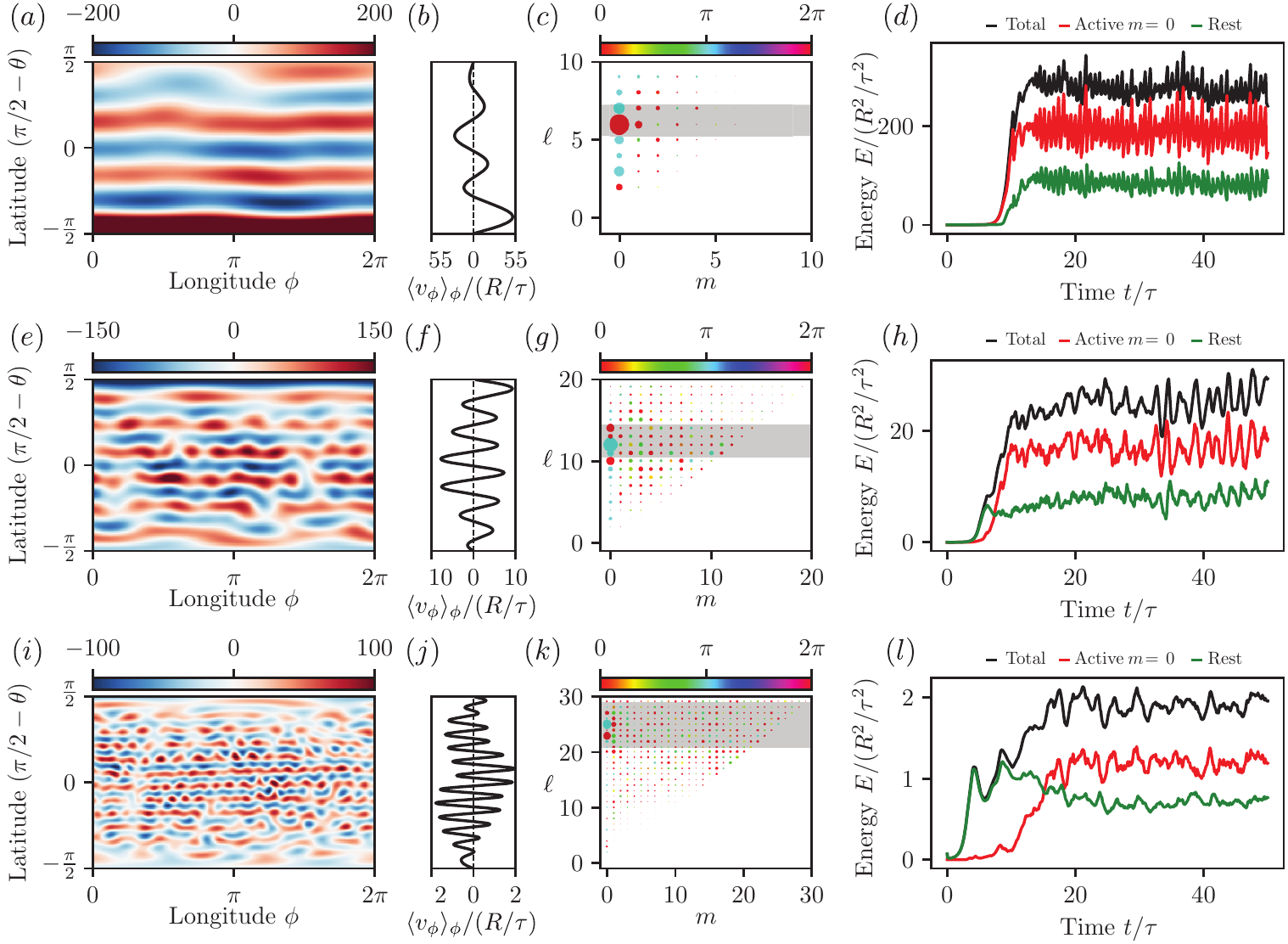}
	\caption{
	Data from steady-state solutions at $t/\tau=15$ for the highest rotation rate $\Omega \tau = 500$.  
	The rows correspond to $R/\Lambda = 2$ $(a\textrm{--}d)$, $4$ $(e\textrm{--}h)$ and $8$ $(i\textrm{--}l)$.
	Panels ($a, e, i$) show Mercator projections of the dimensionless vorticity $\zeta \tau$. 
	Panels ($b, f, j$) show the zonal-mean azimuthal velocities $\langle v_\phi \rangle_\phi/(R/\tau)$.
	Panels ($c, g, k)$ show spherical harmonic decomposition of dimensionless vorticity $\zeta \tau$ with marker size indicating amplitude and color indicating phase.
	All plots indicate the existence of dominant zonal jets with $m=0$ and $\ell$'s within the active band indicated in grey. These modes are close to the exact solutions in figure \ref{Fig:Patterns}($j$--$l$). Panels ($d, h, l$) show time-variation of the energy of all the modes (black), active $m=0$ modes (red), and all other modes (green); the energy contained in the active $m=0$ accounts for most of the total energy in the statistically stationary state. See also supplementary movie 2.}
	\label{Fig:Modes}
\end{figure}

\subsubsection{$\beta$--plane equations}
When the vortical patterns are much smaller than the radius of the sphere ($R/\Lambda \gg 1$), one can linearize  around a reference co-latitude $\theta_0$ to produce a local model. We define metric coordinates in the directions of increasing $\phi$ and decreasing $\theta$, respectively, by $x = R \sin (\theta_0) \phi$ and $y = R(\theta_0 - \theta)$. In these coordinates the dynamical equations are 
\bse
\be
\lap_c \psi &=& -\zeta,\\
\p_t \zeta + J_c (\psi, \zeta) &=& \lap_c F(\lap_c) \zeta +
\beta \p_x \psi,
\label{Eqn:BetaPlane}
\ee
\ese
where $J_c(\psi, \zeta) = \p_y \psi \p_x \zeta - \p_x \psi \p_y \zeta$ \rev{is the Cartesian Jacobian determinant}, $\lap_c$ is  the Cartesian Laplacian and the namesake $\beta$ parameter is given by $\beta = 2 \Omega \sin \theta_0 / R$.
The $\beta-$plane equations preserve the effect of a varying Coriolis parameter $2 \Omega \cos \theta$ while simplifying the spatial operators. 
Rotational effects are accounted for by the term proportional to $\beta$.
\rev{
\subsubsection{Characteristic length and time scales}

The turbulent outer layers of rotating stars and planets ubiquitously contain east-west (zonal) jets of various scales and strengths.
Determining physical processes that generate and maintain these jets is an important problem in planetary science.
A variety of theories have arisen describing jet formation, particularly on the $\beta$-plane, including the arrest of the inverse cascade in rotating turbulence by nonlinear Rossby waves \citep{Rhines:2006cu,Vallis:1993fs}, and as a bifurcation in the statistical dynamics of the zonal flow as a function of the intensity of background homogeneous turbulence \citep{Srinivasan:2012im,Tobias:2013hk}.
These theories predict that jet formation may depend on a variety of length and timescales, such as the Rhines scale $L_R = \sqrt{\mathcal{U}/\beta}$, the scale at which small-scale forcing injecting energy at a rate $\epsilon$ is effected by rotation $L_\epsilon = (\epsilon / \beta^3)^{1/5}$ \citep{Galperin:2010ix}, and the growth rate of unstable perturbations to the zonal flow in statistical models \citep{Bakas:2019hk}.
The GNS model investigated here provides a simplified setting for examining the dynamics of jets by parameterizing the jet-formation physics, rather than attempting to resolve the details of the underlying formation processes. If one is interested in matching the effective GNS parameters to specific length and velocity scales of more detailed models, then guidance can be drawn from the observation that typical zonal jets in the GNS model  have width  $\sim\Lambda$ and r.m.s. velocity $\sim \Lambda\sqrt{\Omega/\tau}$ in the rotation dominated regime $\Omega\tau> 1$.
}

\begin{figure}
    \centering
    \includegraphics[width=0.9\textwidth]{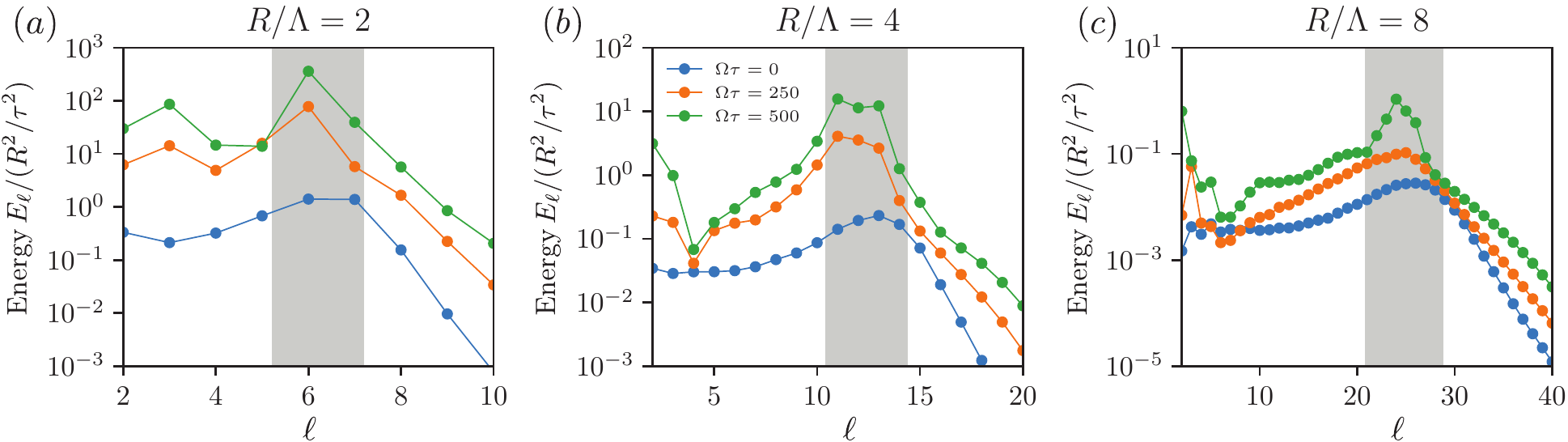}
    \caption{\rev{Energy spectra for $(a) R/\Lambda = 2$, $(b) R/\Lambda = 4$ and $(c) R/\Lambda = 8$. The grey shaded region indicates the active bandwidth where the spectra show a peak. The spectra for $(R/\Lambda, \Omega \tau) = (4, 250), (4, 500)$ and $(8, 500)$ have been obtained from an ensemble average of 10 simulations with random initial conditions.}}
    \label{Fig:EnergySpectra}
\end{figure}

\section{Exact time-dependent solutions}
Exact solutions can be constructed on the sphere as well as on the local $\beta-$plane. Although not stable, these solutions will provide an intuitive understanding of the numerical results in \S \ref{Sec:Simulations}, similar to the role of exact coherent structures \citep{Waleffe2001,Wedin2004} in classical turbulence. 

\subsection{Global solutions}
Exact time-dependent solutions to \eqref{Eqn:GNSRotating} can be constructed as superpositions of normal spherical harmonic modes $Y_\ell^m(\theta, \phi)$ as
\be
\bigg[ \psi(\theta, \phi, t), \zeta(\theta, \phi, t) \bigg] &=& \bigg[1, \ell_\pm(\ell_\pm+1)\bigg] \bigg( \psi_{j} (\theta) + \psi_{w} (\theta, \phi, t) \bigg), 
\label{Eq:SolutionSphere}
\ee
with
\be
\psi_j(\theta) = \mathcal{A}_0 Y_{\ell_\pm}^0(\theta),
\qquad\quad 
\psi_w(\theta, \phi, t) = \Real \left[ \sum_{m=1}^{\ell_\pm} \mathcal{A}_m Y_{\ell_\pm}^m (\theta, \phi) \exp(-\Iota \sigma_m t) \right],
\ee
where $\mathcal{A}_m$ for $m=0,1,...$ are constants,  $\ell_\pm$  are the roots of 
\be
F\bigl( -\ell_\pm(\ell_\pm+1) + 4 \bigr)= 0
\label{Eq:ellRoots}
\ee
and $\sigma_m$ satisfies the dispersion relation 
\be
c_p^\phi = \frac{\sigma_m}{m} = \frac{-2 \Omega}{\ell_\pm (\ell_\pm +1)}.
\label{Eq:DispersionSphere}
\ee
\begin{figure}
    \centering
    \includegraphics[width=0.9\textwidth]{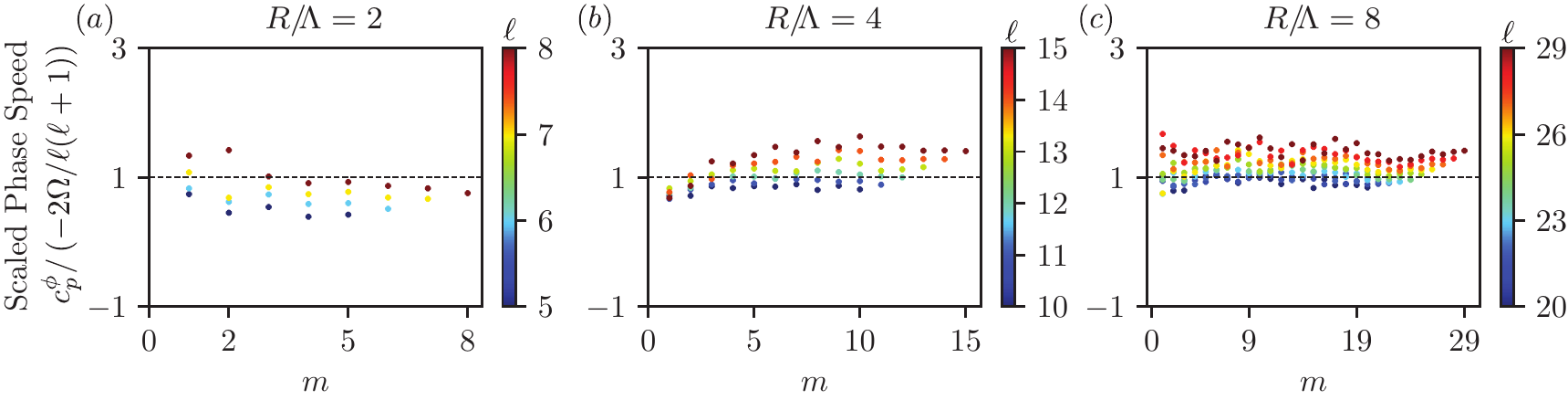}
    \caption{Phase speed of the spherical harmonic modes $(\ell, m)$ in the forcing bandwidth, normalized by the analytical phase speed in \eqref{Eq:DispersionSphere}, for $\Omega \tau = 500$ and different values of $R/\Lambda$. The dotted line indicates the value $1$ for comparison.}
    \label{Fig:Rossby-sphere}
\end{figure}
These solutions to \eqref{Eqn:GNSRotating} are possible because the \rev{Jacobian determinant $J$  vanishes if the stream function is a superposition} of spherical harmonics $Y_{\ell}^m $ with fixed $\ell$. 
We choose $\ell$ to be one of the roots of the polynomial $F$ given in equation  \eqref{Eq:ellRoots}. 
Equation \eqref{Eq:DispersionSphere} describes the dispersion of normal-mode Rossby-Haurwitz waves \citep{Hoskins1973a,Lynch2009,Madden2018}. These  are well-known solutions of the barotropic vorticity equation \citep{Thompson1982}  and  propagate in the direction opposite to the sphere's rotation with phase speed~$c_p^\phi$.
Overall, the exact solutions in \eqref{Eq:SolutionSphere} are a combination of time-independent zonal jets ($\psi_{j}$) and time-varying Rossby-Haurwitz waves ($\psi_w$).  The time-independent zonal jets are spherical harmonics $Y_\ell^0(\theta)$ which consist of alternating crests and troughs. A selection of such modes corresponding to different $\ell$'s are shown in figure \ref{Fig:Patterns}($j$--$l$). 

\subsection{$\beta-$plane solutions}
Similar to the procedure on the full sphere, exact solutions to \eqref{Eqn:BetaPlane} can be constructed by considering superpositions of Fourier modes with wave vectors $\vv{k}$ that correspond to the neutral modes of the pattern forming operator. Hence, the exact solutions are 
\be
\bigg[ \psi(x, y, t), \zeta(x, y, t) \bigg] &=& \bigg[1, k_\pm^2 \bigg] \bigg( \psi_{j} (y) + \psi_{w} (x, y, t) \bigg), 
\label{Eq:SolutionBetaPlane}
\ee
with
\be
\psi_j(y) = \Real \left[ \mathcal{A}_0 \exp (\Iota k_\pm y)\right],\quad
\psi_w(x, y, t) = \Real \left[ \sum_{|\vv{k}| = k_\pm} \mathcal{A}_{\vv{k}} \exp(\Iota (\vv{k} \cdot \vv{x}- \sigma_{\vv{k} } t)) \right],\quad
\ee
where $\mathcal{A}_{\vv{k}}$ are constants, $k_\pm$  are the positive roots  of $F(-k_\pm^2) = 0$,
and $\sigma_{\vv{k}}$ satisfies the Rossby-wave dispersion relation \citep{Pedlosky2003}
\be
c_p^x =\frac{ \sigma_{\vv{k}}}{k_x} = \frac{-\beta k_x}{|\vv{k}|^2}.
\label{Eq:DispersionBetaPlane}
\ee
Here again, solutions are a combination of a time-independent zonal flow ($\psi_j(y)$) and time-varying Rossby waves ($\psi_w(x,y,t)$). For the parameters at which the $\beta-$plane approximation holds, the expression for the phase speed \eqref{Eq:DispersionBetaPlane} provides an explicit dependence on the co-latitude $\theta$ through the parameter $\beta = 2 \Omega \sin \theta_0/R$.
The dynamics at the poles are similar to those on a non-rotating flat plane and the non-inertial effects matter the most at the equator.

\begin{figure}
    \centering
    \includegraphics[width=0.9\textwidth]{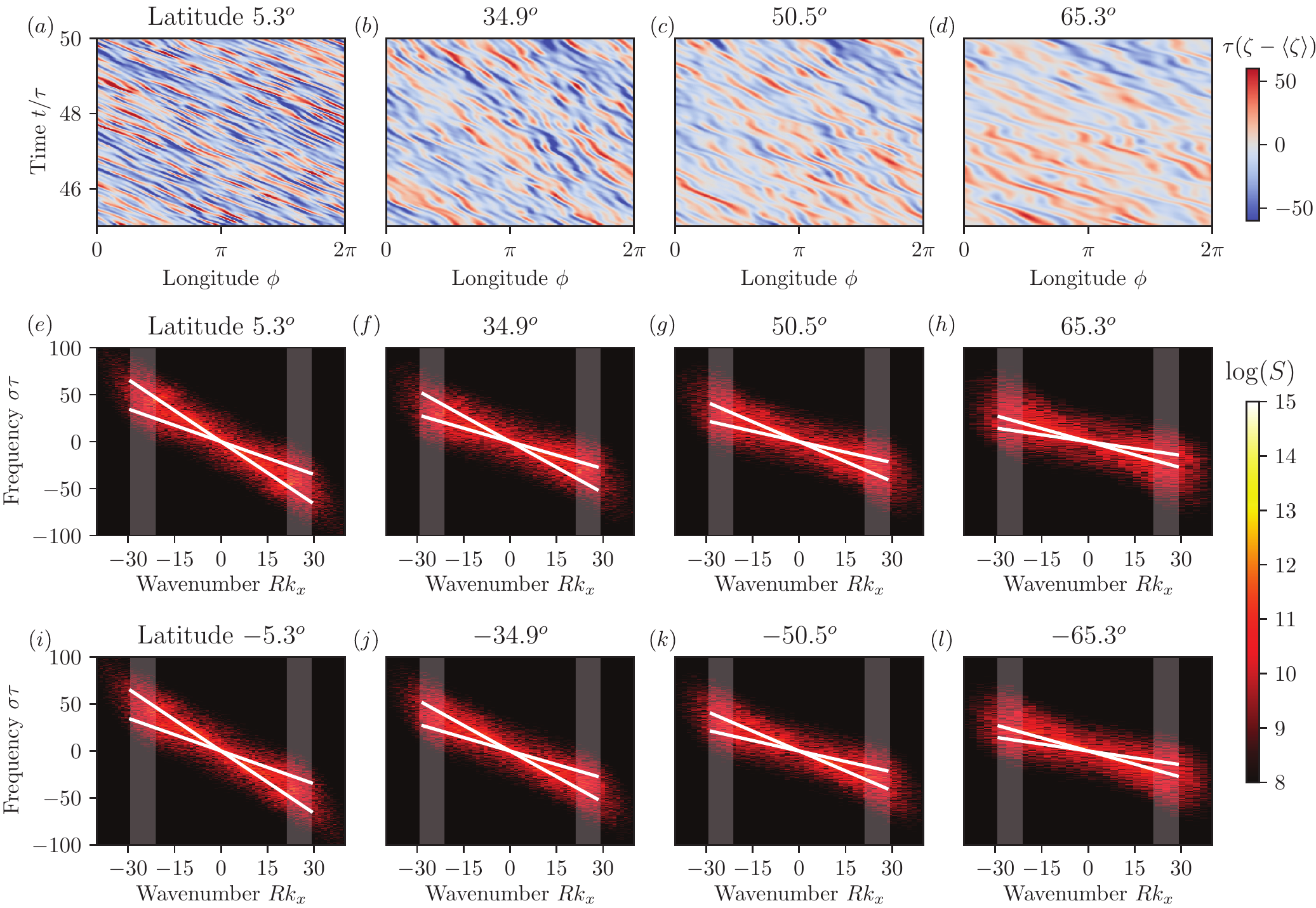}
    \caption{($a$--$d$) Time-space diagrams of the deviation of vorticity,  $\zeta - \langle \zeta \rangle $, where $\langle \cdot \rangle$ is the average over time and space, indicate that the phase speed of the westward propagating Rossby waves in the local $\beta-$plane changes with latitude. ($e$--$l$) Logarithm of the power spectral density, $S = |\tau \hat{\zeta} (k_x, \sigma)|^2$, where $\hat{\zeta}$ is the discrete Fourier transform, at different \rev{northern ($e$--$h$) and southern ($i$--$l$) latitudes.} The grey regions indicate the forcing bandwidth with $k_- < |k_x| < k_+$ justifying the rapid decay of the power spectral density for $|k_x| > k_+$. The white lines in each panel show the analytical dispersion relation from \eqref{Eq:DispersionBetaPlane} with $|\vv{k}| = k_+$ and $|\vv{k}| = k_-$; the one with the steeper slope corresponds to $k_-$.  These predictions  capture the variance of power spectral density. }
    \label{Fig:Rossby-beta}
\end{figure}

\section{Simulations} \label{Sec:Simulations}
%
Direct numerical simulations of \eqref{Eqn:GNSRotating} were performed using a spectral code based on the open-source Dedalus framework \citep{Burns2019}. 
The code uses a pseudo-spectral method with a basis of spin-weighted spherical harmonics \citep{Lecoanet2019a,Vasil2019a}; see appendix of \cite{Mickelin2018}.  A spectral expansion with a cut-off $\ell_{max} = 256$ suffices to obtain converged solutions. The simulations are initialized with a random stream function and evolved from time  $t/\tau=0$ to $t/\tau = 50$. In all simulations, we vary the parameters $R/\Lambda$ and $\Omega \tau$ for fixed dimensionless bandwidth $\kappa \Lambda = 1$. Narrow-band driving with  $\kappa \Lambda \ll 1$ leads to  `burst' dynamics \citep{Mickelin2018} whereas broad-band driving $\kappa \Lambda \gg 1$ leads to classical turbulence \citep{Frisch1995}. The simulations settle onto statistically stationary  flow states after initial relaxation periods during which the active stresses continuously inject energy until the forcing and dissipation balance. The analysis below focuses on the statistically stationary states.

Figures \ref{Fig:Patterns}($a$--$i$) show snapshots of the dimensionless relative vorticity $\zeta \tau$ for a range of $R/\Lambda$ and $\Omega \tau$ at $t/\tau = 15$. In the non-rotating case $\Omega \tau = 0$, we attain solutions akin to those obtained by \cite{Mickelin2018}.  When the dimensionless rotation rate $\Omega \tau$ is increased, the flow becomes zonal, that is, the $\phi-$variation in the vorticity field decreases. At the highest rotation rate $\Omega \tau = 500,$ the vorticity field contains alternating bands of high and low vorticity with a characteristic width. For comparison, we plot the steady state solutions $Y_{\lceil \ell_- \rceil }^0 (\theta)$, where $\lceil \cdot \rceil$ is the ceiling function, in figures \ref{Fig:Patterns} ($j$--$l$). This corresponds to the smallest $\ell$ inside the active band. The formation of zonal flows in our model is consistent with the view \citep{Parker2013, Galperin2019} that such flow structures can be described within a  generic pattern formation framework. 

To better visualize the banded solutions for high rotation rates, we plot Mercator projections of the vorticity in figure \ref{Fig:Modes} (panels $a, e, i$). The banded nature of the vorticity is also reflected in the alternating structure of the mean azimuthal velocity $\langle v_{\phi}\rangle_\phi$ (panels $b, f, j$). The predominant scales in the flow field can be measured using the spherical harmonic decomposition of the relative vorticity. Since $\zeta \tau$ is a real field, we plot only the coefficients with positive $m$ in panels $c, g, k$. The largest modes have $m=0$ with $\ell$ values in the active band of the GNS model (indicated in grey). Panels ($d$--$l$) of figure \ref{Fig:Modes} show the total energy and the energy contained in the active $m=0$ modes as a function of time, calculated from the spherical harmonic coefficients as
\be
\frac{E(t)}{R^2/\tau^2} = \frac{\tau^2}{R^2}\sum_{m, \ell \neq 0} E_{m, \ell} (t) =  \frac{1}{2} \sum_{m, \ell \neq 0} \frac{|\tau \hat{\zeta}_{m, \ell}(t)|^2}{R^2\ell (\ell +1)}.
\ee
After the initial relaxation phase, when the energy injection balances the energy dissipation, the total energy in the system fluctuates around a statistical mean. The active $m=0$ modes carry most of the total energy, implying that the bulk dynamics are dominated by these few modes. This also explains the bandedness of the flow patterns since spherical harmonics with $m=0$  do not vary with the azimuthal angle $ \phi $; see figure \ref{Fig:Patterns} (panels $j$--$l$). \rev{Time-averaged energy spectra of the statistically steady states are plotted in figure \ref{Fig:EnergySpectra}. The energy shows a clear peak within the active bandwidth further suggesting that the active modes carry most of the total energy.}

Strikingly, the statistically stationary states exhibit Rossby waves. For high rotation rates, the Rossby number $Ro$ defined by \eqref{Eqn:RossbyNumber} is $\ll 1$. Thus, we can directly compare the linear phase speed given by \eqref{Eq:DispersionSphere} with the slopes of the linear least squares fits to the phase evolution of the coefficients of vorticity $\hat{\zeta}_{m,\ell}(t)$ from the simulations. The normalized phase speed of the modes for different values of $R/\Lambda$ and $\Omega \tau = 500$ is shown in figure \ref{Fig:Rossby-sphere}. The plotted modes have $\ell$ in the active bandwidth, corresponding to the grey regions in figure \ref{Fig:Modes} (panels $c, g, k$).
The phase speed of the modes from the simulations are close to 1 when normalized by the analytical prediction (figure \ref{Fig:Rossby-sphere}), implying that the linearized theory captures the main characteristics of the nonlinear dynamics.

We also analyze the phase speed of the waves as a function of latitude. According to the dispersion relation in \eqref{Eq:DispersionBetaPlane}, the Rossby-wave phase speed $c_p^x$ depends on the co-latitude $\theta_0$ through $\beta = 2 \Omega \sin \theta_0/R$. To check this prediction, we examine the local dynamics at a number of discrete latitudes ($\pi/2 - \theta_0$) shown in panels ($a$--$d$) of figure  \ref{Fig:Rossby-beta} for $R/\Lambda = 8$ and $\Omega \tau = 500$ (the $\beta-$plane approximation is valid for these parameters). We show the 2D discrete Fourier transform in time and spatial coordinate $x = R(\sin \theta_0) \phi$ in panels ($e$--$l$) for the same \rev{northern and southern} latitudes. The unstable modes lie within the forcing bandwidth or when $k_- < |\vv{k}|<k_+$. Hence, we plot the expected wave dispersion \eqref{Eq:DispersionBetaPlane} making the approximations $|\vv{k}| \simeq k_{+}$ and $|\vv{k}| \simeq k_{-}$. This produces two analytical curves for $\sigma(k_x)$ which are linear in $k_x$ for $k_x < k_+$.
These curves capture the spread of the spectral power in the nonlinear dynamics at every latitude; see the white lines in figures \ref{Fig:Rossby-beta}($e$--$l$). We subsequently infer that the nonlinear, statistically stationary states contain modes with phase speeds matching those of linear $\beta-$plane Rossby waves.

\section{Conclusions} \label{Sec:Conclusion}
We have presented analytical and numerical solutions of generalized Navier-Stokes (GNS) equations on a 2D rotating sphere. This phenomenological model generalizes the widely studied barotropic vorticity equation by adding an internal forcing that injects energy within a fixed spectral bandwidth. We derived a family of exact time-dependent solutions to the GNS equations on the rotating sphere as well as in the local $\beta-$plane. These solutions correspond to a superposition of zonal jets and westward-propagating Rossby waves.  Simulations at high rotation rates confirm that the statistically stationary states are close to these exact solutions. We further showed that the phase speeds of waves in the simulations agree with those predicted for linear Rossby waves. Our results suggest that the GNS framework can serve as a useful minimal model for providing analytical insight into complex flows on rotating spheres, such as planetary atmospheres. It is possible to extend the GNS approach to incorporate more than one dominant length scale by modifying the functional form of the spectral forcing accordingly.

The authors thank Jonasz S\l{}omka, Henrik Ronellenfitsch, Glenn Flierl and Boris Galperin for helpful  discussions. 
The authors acknowledge Geoffrey Vasil and Daniel Lecoanet for the development of the numerical model used in \cite{Mickelin2018}, which also formed the basis of the numerical model in this work.

The authors report no conflict of interest.

\rev{
\appendix
\section{Formulas for $\Lambda, \tau$ and $\kappa$} \label{App:Scales}
For a sphere of radius $R$, 
 $(\Lambda, \tau,\kappa)$ are related to $(\Gamma_0, \Gamma_2,\Gamma_4)$ by \citep{Mickelin2018}:

\bse 
\be
\Lambda &=& \frac{2 \pi R}{2 \sqrt{\frac{17}{4} - \frac{\Gamma_2}{2 \Gamma_4} R^2} - 1},\\
\tau &=& \left[\left(\frac{\Gamma_2}{2 \Gamma_4} - \frac{2}{R^2}\right) \left(\Gamma_0 - \frac{\Gamma_2^2}{4 \Gamma_4}\right)\right]^{-1},  \\
\kappa &=& \left( \frac{17}{2 R^2} - \frac{\Gamma_2}{\Gamma_4} - 2 \sqrt{\frac{17^2}{16 R^4} - \frac{17}{4R^2} \frac{\Gamma_2}{\Gamma_4} + \frac{\Gamma_0}{\Gamma_4} } \right)^{1/2}.
\ee
\label{Eqn:SphereFormulae}
\ese

Letting $R \rightarrow \infty$ in \eqref{Eqn:SphereFormulae}, one obtains for the planar case \citep{Slomka2017a}
\be
\Lambda = \pi \sqrt{\frac{2 \Gamma_4}{- \Gamma_2}}, 
\qquad
\tau = \left[\frac{\Gamma_2}{2 \Gamma_4}  \left(\Gamma_0 - \frac{\Gamma_2^2}{4 \Gamma_4}\right)\right]^{-1}
, \qquad 
\kappa = \left( - \frac{\Gamma_2}{\Gamma_4} - 2 \sqrt{ \frac{\Gamma_0}{\Gamma_4} } \right)^{1/2}.
\ee

\section{Total angular momentum} \label{App:AngularMomentum}
Taking the surface mass density to be 1, the total angular momentum is given by
\begin{align}
M(t) &= \int_0^\pi \int_0^{2\pi} R \sin \theta v_\phi dA = \int_0^\pi \int_0^{2\pi} R^3 \sin^2 \theta v_\phi d\phi d\theta \nonumber \\
&= \int_0^\pi \int_0^{2\pi} R^3 \sin^2 \theta (-\partial_\theta \psi/R) d\phi d\theta = -R^2 \int_0^\pi \int_0^{2\pi}  \sin^2 \theta \partial_\theta \psi d\phi d\theta. 
\end{align}
Applying integration by parts for the $\theta-$integral,
\begin{align}
M(t) &= -R^2 \int_0^{2\pi} \left\{ \int_0^\pi   \sin^2 \theta \partial_\theta \psi d\theta \right\} d\phi \nonumber \\
&=  -R^2 \int_0^{2\pi} \left\{ \left[ \sin^2 \theta \psi \right]_{0}^\pi - \int_0^\pi 2 \sin\theta \cos \theta \psi d\theta \right\} d\phi \nonumber \\
&= 2R^2 \int_0^{2\pi}\int_0^\pi \sin\theta \cos \theta \, \psi\, d\theta  d\phi. 
\end{align}
We may expand the stream function as $\psi(\theta, \phi) = \sum_{\ell, m} \hat{\psi}_\ell^m(t) Y_\ell^m (\theta, \phi)$. The $\phi-$integral survives only for $m=0$ and $Y_\ell^0(\theta, \phi) = P_\ell^0 (\cos \theta)$ where $P$ represents the associated Legendre polynomials. Also realizing that $P_1^0(\cos \theta) = \cos \theta$, we get
\begin{align}
    M(t) = 4 \pi R^2 \sum_\ell \hat{\psi}_\ell^0(t) \left\{ \int_0^\pi P_1^0 (\cos \theta)   P_\ell^0(\cos\theta) \sin \theta d\theta \right\}.
\end{align}
Finally, using the orthogonality relation 
\begin{equation}
\int_0^\pi P_k^m (\cos \theta) P_\ell^m (\cos \theta) \sin \theta d\theta = \frac{2(\ell+m)!}{(2\ell + 1)(\ell - m)!} \delta_{k,\ell},
\end{equation}
we obtain
\begin{equation}
M(t) = \frac{8 \pi}{3} R^2 \hat{\psi}_1^0 (t). 
\end{equation}
Following the results by \cite{Lynch2003}, it can be shown that there is no contribution to $\hat{\psi}^0_1(t)$ from any triad interactions that result from the nonlinear dynamics. 
To see this, we let $\hat{\psi}_{l_\gamma}^{m_\gamma}(t) = \hat{\psi}_{1}^{0}(t)$ which interacts with coefficients $\hat{\psi}_{\ell_\alpha}^{m_\alpha}$ and $\hat{\psi}_{\ell_\beta}^{m_\beta}$. For a non-vanishing triad interaction, the necessary conditions, $|\ell_\alpha - \ell_\beta| < \ell_\gamma = 1$ and $\ell_\alpha \neq \ell_\beta$, cannot be simultaneously satisfied for $\ell_\gamma = 1$. Additionally, equation \eqref{Eqn:GrowthRateSphere} shows that the GNS forcing is zero for $\ell = 1 $. Thus, $\hat{\psi}^0_1(t)$ remains constant and the total angular momentum is conserved. 
}
\bibliographystyle{jfm}
\bibliography{active-sphere-ref}

\end{document}